\documentclass[reprint, prl, superscriptaddress]{revtex4-1}

\usepackage{graphicx}
\usepackage{color}
\usepackage{amsmath}
\usepackage{bm}

\begin{document}

\title{Distillation of squeezing using a pulsed engineered parametric down-conversion source}

\author{Thomas Dirmeier}
\affiliation{Max Planck Institute for the Science of Light, Staudtstr. 2, 91058 Erlangen, Germany}
\affiliation{Friedrich-Alexander-Universit\"at Erlangen-N\"urnberg (FAU), Department of Physics, Staudtstr. 7/B2, 91058 Erlangen, Germany}
\author{Johannes Tiedau}
\affiliation{Integrated Quantum Optics, Paderborn University,Warburger Str. 100, 33098 Paderborn, Germany}
\author{Imran Khan}
\affiliation{Max Planck Institute for the Science of Light, Staudtstr. 2, 91058 Erlangen, Germany}
\affiliation{Friedrich-Alexander-Universit\"at Erlangen-N\"urnberg (FAU), Department of Physics, Staudtstr. 7/B2, 91058 Erlangen, Germany}
\author{Vahid Ansari}
\affiliation{Integrated Quantum Optics, Paderborn University,Warburger Str. 100, 33098 Paderborn, Germany}
\author{Christian R. M\"uller}
\affiliation{Max Planck Institute for the Science of Light, Staudtstr. 2, 91058 Erlangen, Germany}
\affiliation{Friedrich-Alexander-Universit\"at Erlangen-N\"urnberg (FAU), Department of Physics, Staudtstr. 7/B2, 91058 Erlangen, Germany}
\author{Christine Silberhorn}
\affiliation{Integrated Quantum Optics, Paderborn University,Warburger Str. 100, 33098 Paderborn, Germany}
\author{Christoph Marquardt}
\affiliation{Max Planck Institute for the Science of Light, Staudtstr. 2, 91058 Erlangen, Germany}
\affiliation{Friedrich-Alexander-Universit\"at Erlangen-N\"urnberg (FAU), Department of Physics, Staudtstr. 7/B2, 91058 Erlangen, Germany}
\author{Gerd Leuchs}
\affiliation{Max Planck Institute for the Science of Light, Staudtstr. 2, 91058 Erlangen, Germany}
\affiliation{Friedrich-Alexander-Universit\"at Erlangen-N\"urnberg (FAU), Department of Physics, Staudtstr. 7/B2, 91058 Erlangen, Germany}



\begin{abstract}
Hybrid quantum information processing combines the advantages of discrete and continues variable protocols by realizing protocols consisting of photon counting and homodyne measurements. However, the mode structure of pulsed sources and the properties of the detection schemes often require the use optical filters in order to combine both detection methods in a common experiment. This limits the efficiency and the overall achievable squeezing of the experiment.
In our work, we use photon subtraction to implement the distillation of pulsed squeezed states originating from a genuinely spatially and temporally single-mode parametric down-conversion source in non-linear waveguides. Due to the distillation, we witness an improvement of 0.17~dB from an initial squeezing value of $-1.648 \pm 0.002$~dB, while achieving a purity of 0.58, and confirm the non-Gaussianity of the distilled state via the higher-order cumulants. With this, we demonstrate the source's suitability for scalable hybrid quantum network applications with pulsed quantum light.
\end{abstract}
\maketitle
\section{Introduction}
For practical quantum information processing (QIP), e.g. long-distance quantum communication, the ability to distribute quantum states is essential. However, this ability is limited by decoherence processes during transmission and the  inability to deterministically amplify quantum states. The distillation of quantum resources, such as squeezing or entanglement, provides the means to undo parts of the detrimental effects of transmission \cite{Bennett96}.
\par
In continuous-variable quantum information processing (CV-QIP), the field quadratures of Gaussian states are used to carry the information. Gaussian resource states such as squeezed or two-mode squeezed (entangled) states are routinely generated in laboratories e.g. through parametric down-conversion (PDC) in $\chi^{(2)}$ crystals; either embedded in an optical cavity as an optical parametric oscillator below threshold \cite{Vahlbruch2016,Takanashi2019,Foertsch2015, Otterpohl19} or in a high efficiency single-pass configuration \cite{Eckstein11, Georg13, Yoshino2007}. Pulsed sources offer distinct benefits in terms of energy efficiency and source synchronization. However, they typically emit an ensemble of spatial and temporal modes. The disadvantage of this intrinsic multi-mode nature is that the squeezing is spread over a great number of modes, limiting the overall achievable squeezing in a single one of them, and that the separate detection of different modes requires mode-resolving detection schemes, e.g homodyne detection. In homodyne detection, a signal beam is interfered with a bright reference beam, the local oscillator (LO), and then usually detected by a pair of PIN photo diodes. The detection signal of interest is then retrieved when looking at the difference current of both diodes. Due to this interference, only those components of the signal will be detected which have a non-zero overlap with the LO field.
\par
In addition to the properties of the source that generates the Gaussian states, we have to consider that the distillation of Gaussian states is a challenging task, which requires non-Gaussian (nG) operations \cite{Eisert2002}. This is particularly true for pulsed quantum light, which typically exhibits a rich temporal mode structure \cite{PhysRevX.5.041017,Ra2019}. Among various methods of implementing nG operations \cite{Heersink2006, Dong2008, Hage2008, Ulanov2015}, hybrid discrete- and continuous-variable operations are especially interesting because they combine the distinct features of homodyne and photon counting measurements \cite{Andersen2015, Morin2014, Jeong2014, Masada2015, Ulanov2017, Agudelo2017, Wenger2004,Kurochkin2014, Takahashi2010}. In particular, photon subtraction has been used in many QIP scenarios. The photon subtraction is implemented by detecting a single photon from a pulse after a weakly reflecting beam splitter. Conventional single-photon detectors, in contrast to homodyne detectors, cannot resolve the temporal mode structure of the impinging ultrafast pulsed light field. Due to this mode-insensitivity, the photons are subtracted from an incoherent mixture of all excited modes degrading the purity of the final state \cite{Andi2014}; posing this scheme unsuitable for multi-mode states. Usually, this issue is dealt with by employing a set of spatial and spectral filters \cite{Wenger2004,Kurochkin2014, Takahashi2010}, which in return reduces the success rate of the experiment. As an alternative solution, photon counting detectors can achieve mode-resolution when combined with an additional non-linear process that is mode-selective. Such experiments have been performed but come at the cost of an additional experimental overhead \cite{Ra2017, Ansari2017, Ra2019}.
\par
In our work, we present the successful distillation of pulsed squeezed states using a periodically poled potassium titanyl phosphate (PPKTP) waveguide as engineered source of broadband squeezed states. By using a single-mode waveguiding structure and engineering the non-linear interaction inside the crystal with ultrafast pump pulses, we are able to generate spatially single-mode and spectrally separable pulsed squeezed states of light thus reducing their effective number of modes. This reduced mode-number enables us to demonstrate the efficient combination of photon counting and homodyne measurements in a single protocol while, at the same time, avoiding the use of strict spectral filters. We witness an increase in the detected squeezing due to the distillation and can furthermore verify the non-Gaussian nature of the distilled pulsed quantum state, which both match with the predictions from our theoretical model. We want to emphasize, that the preparation of pulsed quantum states is of particular importance for synchronizing multiple conditional state preparations in network applications. Our results highlight the advantages of pulsed-state source engineering, which consist of a reduction in the experimental complexity and an increase in the efficiency of the state generation, in the context of hybrid QIP.
\section{Theoretical Background}
In hybrid QIP experiments, we aim to combine the features of DV and CV measurements in order to realize operations inaccessible by DV or CV methods alone. A standard hybrid QIP protocol is the distillation of two-mode squeezing via the subtraction of photons from both modes of a two-mode squeezed state. In the following, we model a pulsed two-mode squeezed state generated by our source in an idealized setting. For this, we start by considering a PDC source generating pulsed two-mode squeezed states in a single spatio-temporal mode 
\begin{eqnarray}
	|\psi_{\mathrm{sq}} \rangle = \sqrt{1 - \lambda^2} \sum_{n=0}^{\infty}{\lambda^n | n\rangle| n\rangle}.
	\label{eq:purePDC}
\end{eqnarray}
where the photon number \mbox{$|n\rangle = \frac{1}{\sqrt{n!}} \ \left(\hat{A}^{\dagger}\right)^{n} |0\rangle$} in a specific temporal mode defined by the creation operator \mbox{$\hat{A}^{\dagger} = \int d\omega f(\omega) \hat{a}^{\dagger}(\omega)$} and the generalized squeezing parameter \mbox{$\lambda = \tanh(B)$}, whereby the parameter B describes the parametric gain of the process.
\par
The distillation is then realized by splitting the two-mode squeezed state on beam splitters with identical transmissivity $T$. The reflected portion is sent to click detectors implementing the photon subtraction while the transmitted part is forwarded to the next QIP step. We want to emphasize here that the click detectors are fully, i.e. temporally and spatially, mode-resolving in the sense that they always detect the same spatio-temporal mode that the following processing step operates on. The state after the photon subtraction can then analytically be expressed as \cite{Tim15}
\begin{align}
|\psi_{\mathrm{sub(2)}}\rangle \propto \sqrt{1-\lambda^2}\sum_{n = 0}^{\infty} \lambda^{n+1}(n+1)T^{2n}(1-T^2)|n\rangle|n\rangle.
\label{eq:SMdist}
\end{align}
The photon subtraction grants a larger weight to photon number components of higher order and increases the degree of squeezing of the state. In order to reduce the experimental complexity, we interfere both modes of the photon-subtracted state in Eq.~(\ref{eq:SMdist}) and record the quadrature data of one of the output ports. In its density matrix representation, the detected state can be calculated via the expression
\begin{align}
    \hat{\rho}_{\mathrm{sub}(2)} = \mathrm{tr}_A(\hat{U}_{A,B}^{\dagger}|\psi_{\mathrm{sub(2)}}\rangle\langle\psi_{\mathrm{sub(2)}}|\hat{U}_{A,B}),
    \label{eq:finalstate}
\end{align}
with the partial trace $\mathrm{tr}_A$ indicating that one of the modes is not used for measurements. The operator $\hat{U}_{A,B}$ describes the symmetric beam splitter on which both modes are interfered and the indices A and B denote its output ports. For our simulations, we calculate the state in Eq.~(\ref{eq:finalstate}) numerically since it has, to our knowledge, no closed analytic representation.
\par
In reality however, pulsed PDC processes emit temporal multi-mode two-mode squeezed states. The broadband generation process is usually described in the Heisenberg picture as
\begin{align}
|\psi \rangle = \exp[-\frac{i}{\hbar} B \int \!\mathrm{d}\omega_s \mathrm{d}\omega_i f(\omega_s,\omega_i) \hat{a}^{\dagger}_s(\omega_s)\hat{b}^{\dagger}_i(\omega_i) + h.c.]|0\rangle.
\end{align}
Here, the joint spectral amplitude (JSA) function \mbox{$f(\omega_s,\omega_i) = \alpha(\omega_s,\omega_i) \Phi(\omega_s,\omega_i)$}, which is the product from phase matching function and the pump field envelope, describes the range of allowed signal and idler frequencies in the PDC process at hand. In general, the JSA can be decomposed in a set of mutually orthogonal modes using the Schmidt decomposition \mbox{$f(\omega_s,\omega_i) = \sum_k{c_k g_k(\omega_s)h_k(\omega_i)}$} with Schmidt coefficients $c_k$, which obey the normalization condition \mbox{$\sum_k{c_k^2 = 1}$} \cite{Christ2011}. The effective number $K$ of modes in a such PDC process are equally given by the Schmidt coefficient, it can be expressed via \mbox{$K = \sum_k 1/|c_k|^4$}. As such, the state generated in a PDC process can be rewritten in terms of the Schmidt decomposition which results in general in a product state of a multitude of individually two-mode squeezed beams, one in each Schmidt mode. Mathematically, this is represented as
\begin{align}
|\psi \rangle  &= \exp[-\frac{i}{\hbar} B \int \!\mathrm{d}\omega_s \mathrm{d}\omega_i \sum_k{ c_k \hat{A}_{k,s}^{\dagger}\hat{A}_{k,i}^{\dagger}+ h.c.}]|0\rangle \nonumber\\
&= \bigotimes_{{k}}{\sqrt{1 - \lambda_{{k}}^2} \sum_{n_{{k}}=0}^{\infty}{\lambda_{{k}}^n | n_{{k}}\rangle| n_{{k}}\rangle}}.
\end{align}
with the temporal mode creation operators \mbox{$\hat{A}_{k} = \int \!\mathrm{d}\omega \ \hat{a}_k(\omega)f_k(\omega)$} and a mode-dependent squeezing strength \mbox{$\lambda_{{k}} = \tanh(c_{{k}}B)$}. It has been shown that it is possible to use dispersion engineering~\cite{laserphysics2005} in order to generate pulsed two-mode squeezed states with a factorable joint spectral amplitude function. This corresponds to the generation of quantum states in a single spatio-temporal mode and has been demonstrated in previous DV-QIP experiments by preparing and interfering high-purity single photons \cite{Eckstein11, Georg13, Georg16, Francis-Jones16, Bruno14, Kaneda16}.
\par
Yet in the actual experiment, the generated quantum states still exhibit small contributions from higher-order Schmidt modes, which need to be accounted for in the squeezing distillation process, if not dealt with by using optical filters. This is due to the fact that typical click detectors behave not ideally when detecting a multi-mode light field. In our squeezing distillation experiment as shown in Fig.~\ref{figexperiment}, we need to consider the possibility that the avalanche photo diodes (APDs) randomly detect photons from different Schmidt modes. In our case however, we can restrict ourselves to the first two orders of Schmidt modes as the influence of modes of order higher than the first two is vanishing.
\par
When one of the click detectors detects a photon, it will randomly fire from a photon from one of the Schmidt modes. We are triggering our homodyne measurement on the registration of a coincidence event, consequently several different realizations of that coincidence lead to a homodyne measurement in the multi-mode case. A coincidence might stem from two photons of the Schmidt mode coinciding with the LO, a photon in the LO mode and any other mode, and two photons from any other mode that we cannot distinguish between. Therefore, we model the detected state as a mixed state consisting of the superposition of the intended distilled state $\hat{\rho}_{\mathrm{sub(2)}}$, a state with a photon subtracted in one of the two modes $\hat{\rho}_{\mathrm{sub(1)}}$, and an undistilled squeezed state $\hat{\rho}_{\mathrm{sq}}$, each weighted with a parameter $\alpha_i$
\begin{eqnarray}\label{eq:state}
	\hat{\rho} = \alpha_0 \hat{\rho}_{\mathrm{sub(2)}} + \alpha_1\hat{\rho}_{\mathrm{sub(1)}} + \alpha_2\hat{\rho}_{\mathrm{sq}}.
\end{eqnarray}
To calculate the parameters $\alpha_{\mathrm{i}}$, we need to calculate the different detection probabilities in each Schmidt mode $k$ via \mbox{$p^{(k)}_{mn} = \mathrm{tr}\left(|m,n\rangle\langle m,n|_k \hat{\rho}^{(k)}_{\mathrm{sq}} \right)$}, $m$ and $n$ represent here the photon number subtracted from the signal and idler modes. Due to the low mean photon numbers in the tap-off arms, we limit ourselves to either subtracting a single or no photon at a time. We model the limited heralding efficiency $\eta$ of our experiment using a Binomial distribution \mbox{$L_{\mathrm{i,j}} = \sum_{\mathrm{j}}{\binom{j}{i} \eta^i (1-\eta)^{j-i}}$} and use it to rescale the detection probabilities via the relation $p'^{(k)}_{m'n'} = \sum_{m,n}{ L_{m'm}  p^{(k)}_{mn}  L_{n'n}}$. The weighting parameters $\alpha_{\mathrm{i}}$ are then by given by
\begin{align}
\alpha_0 = \frac{p'^{(0)}_{11}  p'^{(1)}_{00}}{p'^{(0)}_{11}  p'^{(1)}_{00} + p'^{(0)}_{00}  p'^{(1)}_{11} + 2p'^{(0)}_{10}  p'^{(1)}_{01}},\nonumber\\ \alpha_1 = \frac{2p'^{(0)}_{10}  p'^{(1)}_{01}}{p'^{(0)}_{11}  p'^{(1)}_{00} + p'^{(0)}_{00}  p'^{(1)}_{11} + 2p'^{(0)}_{10}  p'^{(1)}_{01}} ,\\ \alpha_2 = \frac{p'^{(0)}_{00}  p'^{(1)}_{11}}{p'^{(0)}_{11}  p'^{(1)}_{00} + p'^{(0)}_{00}  p'^{(1)}_{11} + 2p'^{(0)}_{10}  p'^{(1)}_{01}}\nonumber.
\end{align}
The coefficients $\alpha_i$ are directly connected to the mode purity given by the Schmidt coefficients. The more higher-order mode contribution there is to the state, the higher the contribution of non-optimal coincidence events becomes. Both of these ``unwanted'' events show a smaller degree of squeezing than the desired state which results from the spectral mode coinciding with LO mode. This mixed state exhibits, using parameters taken from the characterization of our experiment, a squeezing of -1.89~dB relative to shotnoise which is notably reduced when compared to the ideal case of a pure initial state that shows about -2~dB of squeezing. We want to emphasize that mode impurity, in absence of either filtering or source engineering, limits the amount of achievable squeezing improvement even before losses are considered.

\section{Experimental results}
\subsection{Experimental setup}
\begin{figure}[t]
	\centering
		\includegraphics[width = 0.9\columnwidth]{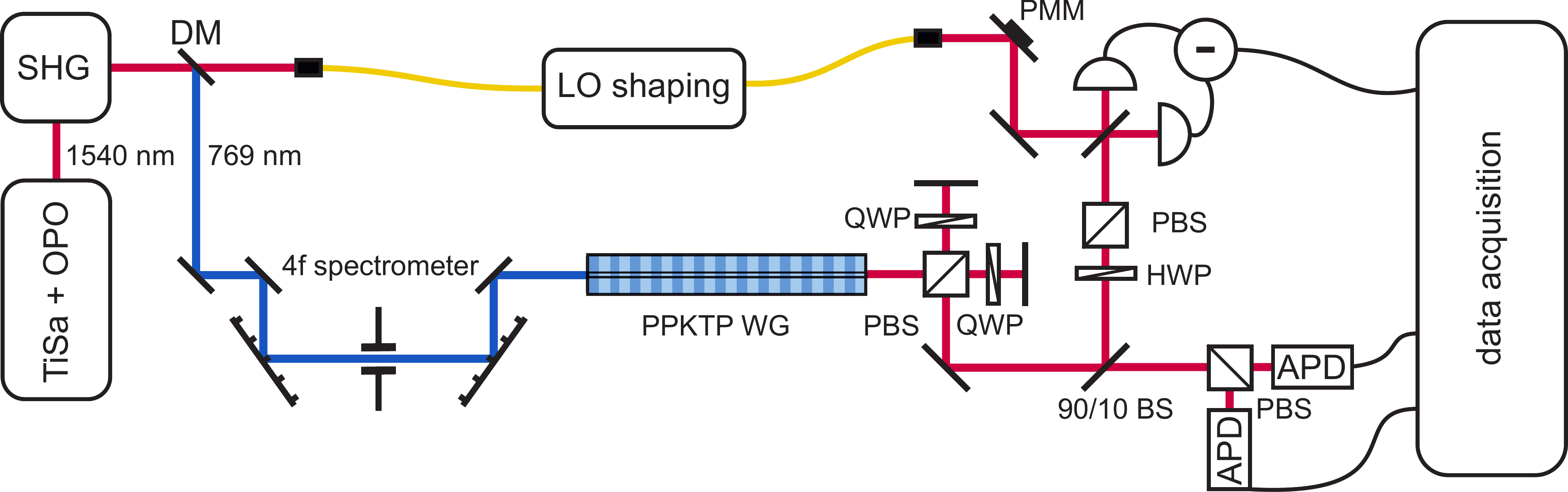}	
\caption{Schematic drawing of the experimental setup: the master optical parametric oscillator (OPO) is frequency doubled to generate the pump pulses for the PDC process. The unconverted portion is separated with a dichroic mirror (DM) and serves as the LO. A 4f spectrometer allows us to choose the pump bandwidth such to achieve spectral decorrelation. After the waveguide, we compensate for the birefringence of KTP before tapping of 10\% of the signal and idler beams for the photon subtraction. The remaining portions are interfered on the combination of a halfwave plate (HWP) and a polarizing beam splitter (PBS) to generate a pulsed, genuinely single mode squeezed beam which is then detected with a homodyne receiver. A piezo-mounted mirror (PMM) allows for continuous scanning the quadrature angle under which we detect the state.}
\label{figexperiment}
\end{figure}
Our experimental setup is described in Fig.~\ref{figexperiment}, a mode-locked femto-second laser at a repetition rate of 82~MHz is pumping a synchronously pumped optical parametric oscillator with its central wavelength at 1540~nm which is used to drive our experiment. The light is frequency doubled via a periodically poled lithium niobate crystal, generating the pump for the PDC process. The up-converted beam is then filtered via a 4f-spectrometer in order to ascertain the spectral decorrelation of the pulsed squeezed states via an optimized pump spectral width \cite{Grice2001, laserphysics2005}. In KTP, a type-II phasematched PDC process can be used to meet this decorrelation condition that allows us to generate genuinely single-mode states. In order to achieve single TM emission, we choose the pump pulse bandwidth to be 2.3~nm and its central wavelength to 769.6~nm so that we achieve the generation of frequency degenerate signal and idler beams around 1540~nm.
\par
After the waveguide, we compensate for the birefringence of PPKTP using a Michelson-like interferometer with adequately chosen arm length so that signal and idler pulses have a high temporal overlap. We realize the photon subtraction by tapping off a small portion of the two-mode squeezed pulses via a 90/10 beam splitter which allows to compromise between a high distillation gain and the data acquisition rate. After the beam splitter, we filter signal and idler pulses using a bandpass with 10~nm FWHM to suppress any background modes. This filter bandwidth is chosen such that it does not influence the spectra of signal and idler that show spectral bandwidths $\Delta\lambda_s = 4.1~\mathrm{nm}$ and $\Delta\lambda_i = 5.8~\mathrm{nm}$ respectively. Thereafter, the modes are split using a polarizing beam splitter and subsequently sent onto a pair of APDs. Coincidence signals from the APDs will then be forwarded to herald a successful photon subtraction in the signal and idler modes. Following the brighter port of the asymmetric beam splitter, signal and idler pulses are interfered with each other on the combination of a half-wave plate and another polarizing beam splitter to generate a pair of single-mode squeezed beams of which one is sent onto a homodyne receiver with adapted LO pulses in order to record the quadrature data. The mixing of both modes of a two-mode squeezed state here can be understood as the inversion of generating two-mode squeezing by interfering two single-mode squeezed beam on a symmetric beam splitter \cite{Christine2001}.
\par
We record the measurement data with a raw distillation rate is about $10~\mathrm{s}^{-1}$, due to technical limitations of the recording hardware however, we are limited to an effective distillation rate of $250~\mathrm{minute}^{-1}$. In total, we record 250000 time traces with a span of 100~$\mathrm{\mu s}$ and a sampling rate of $10^9~\mathrm{Samples}~\mathrm{s}^{-1}$. Each trace contains a single photon subtraction event surrounded by about 8000 squeezed states not subject to the photon subtraction. We assume, that the relative phase between squeezed light and LO stays constant during this time interval. This allows us to reconstruct the relative phase for every trace by using the variance of the squeezed states as a reference and, furthermore, allows us to directly compare the improvement in the quadrature variance due to the squeezing distillation. Further details on the phase estimation routine can be found in appendix~\ref{app}.
\subsection{Characterization}
We characterize the properties of our source by a number of DV measurement techniques using APDs. A Hong-Ou-Mandel interference measurement yields us a visibility of \mbox{$0.75 \pm 0.01$} when interfering the signal and idler beams. The mean photon number $\langle n \rangle$ is estimated to be \mbox{$0.56 \pm 0.001$} photons per pulse which is equivalent to a two-mode squeezing of \mbox{6~dB} at a pump pulse energy of \mbox{$E_p = 41~\mathrm{pJ}$}. The heralding efficiency after the asymmetric beam splitter is \mbox{$\eta = 0.002$}. Finally, we estimate the number of effective time-frequency modes of our squeezed state via a marginal $g^{(2)}$ measurement \cite{Christ2011}, resulting in a value of \mbox{$g^{(2)}= 1.81 \pm 0.05$} and thus corresponding to an effective mode-number of 1.23.
\par
By seeding the PDC process with coherent pulses at \mbox{1540~nm} and measuring the optical power at the detection stage, we estimate the efficiency of the optical elements of the squeezing path to be 0.87. The optimization of the homodyne receiver, mainly the mode overlap between the pulsed LO and the squeezed state, is achieved by using the visibility between the seeded PDC and an attenuated LO as figure of merit. We optimize the visibility by shaping the spectral amplitude and phase of the LO pulses utilizing a commercially available waveshaper (Finisar waveshaper 4000s) and achieve a visibility of up to 0.91 between the seeded PDC and the LO field. When taking the quantum efficiency of 0.90 of the home-built homodyne detector into account, we result at a total receiver efficiency of about 0.72.
\begin{figure}[t]
\centering
	\includegraphics[width = \columnwidth]{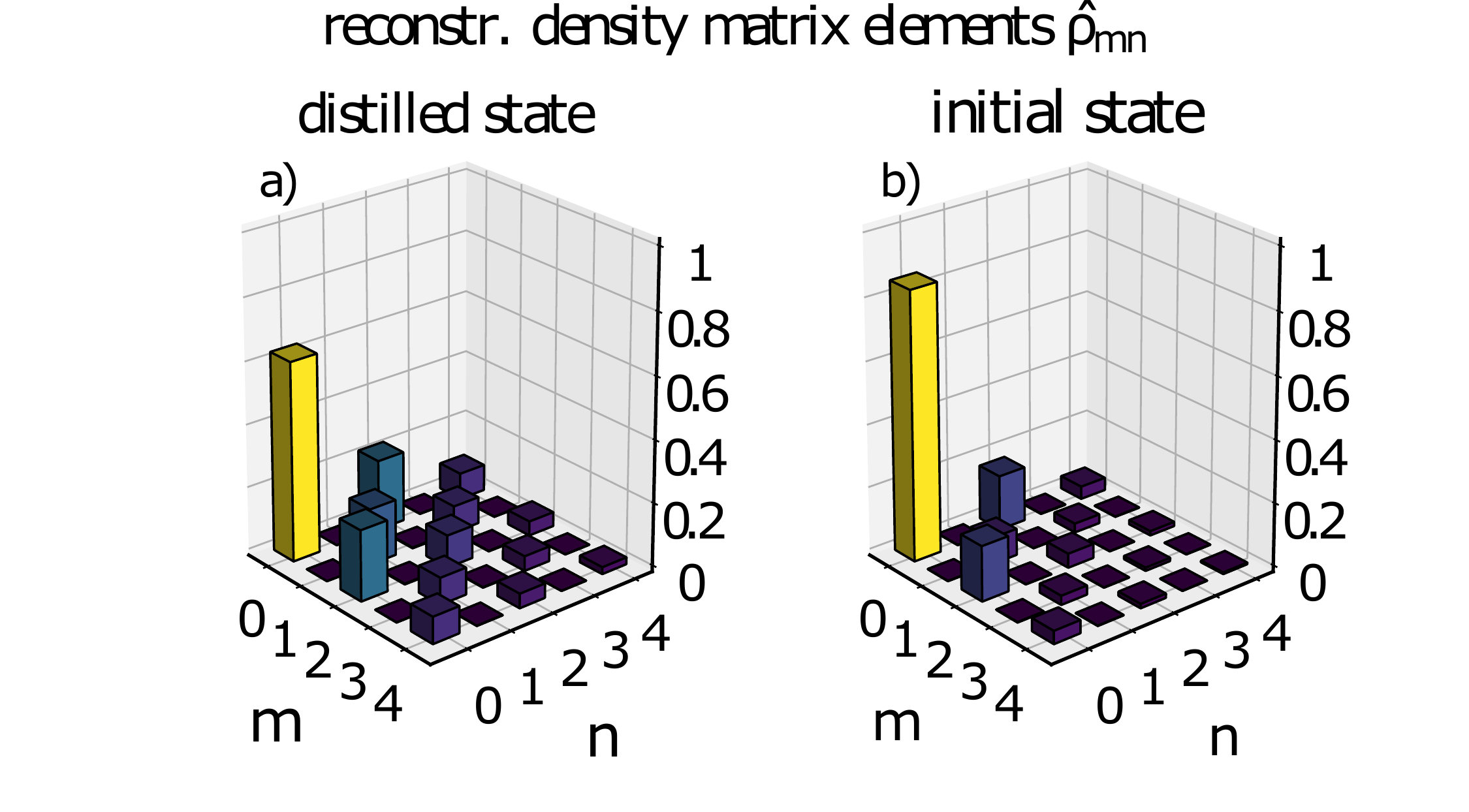}
\caption{The first few elements of the reconstructed density matrices underlying the Wigner functions in Fig.~\ref{fig:wignerfkt}. The distilled state, shown in a), exhibits more pronounced higher-order photon components as is expected for this state when compared to the reconstructed initial state sown in b). The density matrices have been reconstructed from the measurement data and not corrected for detection imperfections.}
\label{fig:densitymatrices}
\end{figure}
\subsection{Squeezing improvement}
\begin{figure}[h]
\centering
	\includegraphics[width = \columnwidth]{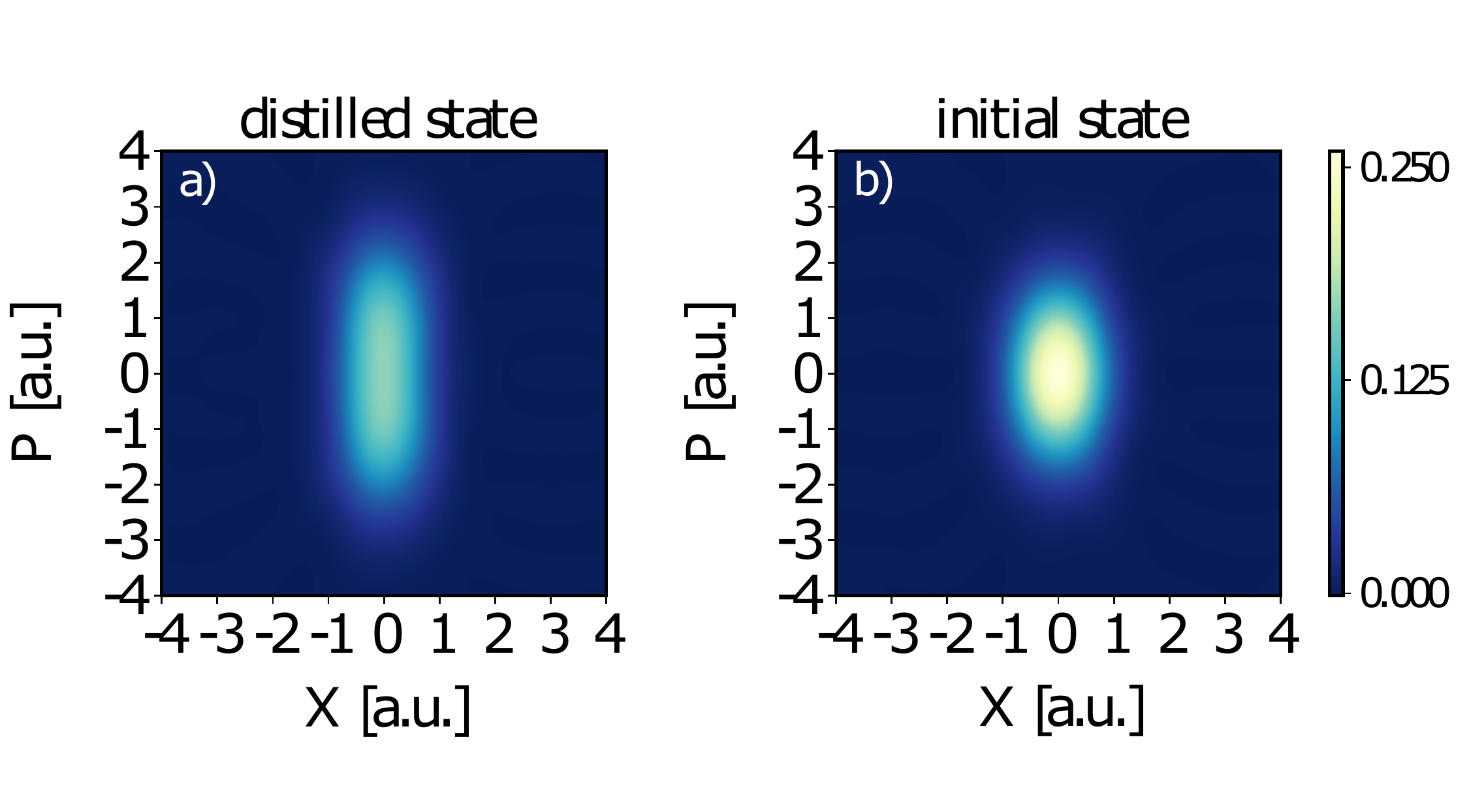}
	\caption{\label{fig:wignerfkt}Contour plots of the Wigner functions of the distilled state a) and the initial state b). The Wigner function of the distilled state is strongly deformed along the P quadrature when compared to the initial state which is due to the effects of the distillation procedure.}
\end{figure}
After successfully subtracting photons from both modes, the higher photon number components of even order become more dominant for the distilled states which is equivalent to an increase in the squeezing. We verify this by comparing the squeezing and anti-squeezing values of the initial and distilled quantum states. To do so, we use the 8000 undistilled pulses surrounding each photon subtraction event to serve as a phase reference to sort the data by their relative phase. This results in a squeezing variance of \mbox{$-1.648 \pm 0.002$~dB} relative to the shot noise for the undistilled state which fits well to the degradation of a 6 dB squeezed state after experiencing the losses of our setup. Due to the distillation step, this variance is increased to \mbox{$-1.82 \pm 0.16$~dB} below shot-noise. Note that these values have been calculated from the sorted raw data. Therefore, the statistics for the distilled events is limited thus increasing the standard deviation when compared to the undistilled squeezed state for which a greater number of states has been sampled.
\par
Reconstructing the density matrices of both states using a maximum likelihood algorithm~\cite{Rehacek2007}, we can directly see this improvement of the quadrature variance in the values of the density matrix elements as shown in Fig.~\ref{fig:densitymatrices}. The increased contribution of the higher-order photon number components can also be visualised by plotting the Wigner functions in Fig.~\ref{fig:wignerfkt}. The distilled state deviates from an elliptical shape, instead it is compressed in the $\hat{X}$ quadrature, representing the higher squeezing, and strongly elongated along the $\hat{P}$ quadrature. In order to understand these findings, we simulate our experimental data with our multi-mode theoretical model discussed earlier. These results are summed up in Table~\ref{tab:results} and show a comparison of the reconstructed initial state $\hat{\rho}_{\mathrm{sq}}$ and the distilled state $\hat{\rho}_{\mathrm{dist}}$ with the simulation results $\hat{\rho}_{\mathrm{sq,sim}}$ and $\hat{\rho}_{\mathrm{dist,sim}}$. The figure of merit to compare theory and experiment is the fidelity between the measured and calculated density matrices, which shows a high agreement between experimental data and theoretical model.
\begin{table}[t]\centering
\caption{\label{tab:results}Comparison of the quadrature variances and the purity of the reconstructed density matrices between the measured data and the results of multi-mode simulations}
  \begin{tabular}{ccccc}
		\hline
& $\hat{\rho}_{\mathrm{sq}}$& $\hat{\rho}_{\mathrm{sq,sim}}$ & $\hat{\rho}_{\mathrm{dist}}$& $\hat{\rho}_{\mathrm{dist,sim}}$\\
     \hline
Var($\hat{X}) [\mathrm{dB}]$& $-1.648 \pm 0.002$& $-1.67$&$-1.82 \pm 0.16$&$ -1.89$\\
Var($\hat{P}) [\mathrm{dB}]$& $3.456 \pm 0.002$& 3.38& $6.28 \pm 0.15$& 6.06\\
tr($\hat{\rho}^2$)& 0.80&0.82 & 0.58& 0.61\\
\hline
Fidelity&\multicolumn{2}{c}{99.9 \%}&\multicolumn{2}{c}{99.8 \%} \\
    \hline
  \end{tabular}
\end{table}
We want to emphasize here that the purity $\mathrm{tr}(\hat{\rho}^2)$ of the squeezed state can solely be explained by the optical losses present in the system. As a result of this, we are also able to see the additional drop in the purity value for the distilled state resulting from the non-Gaussian distillation process.
\subsection{Verifying the non-Gaussianity}
\begin{figure}[t]
\centering
      \includegraphics[width=\columnwidth]{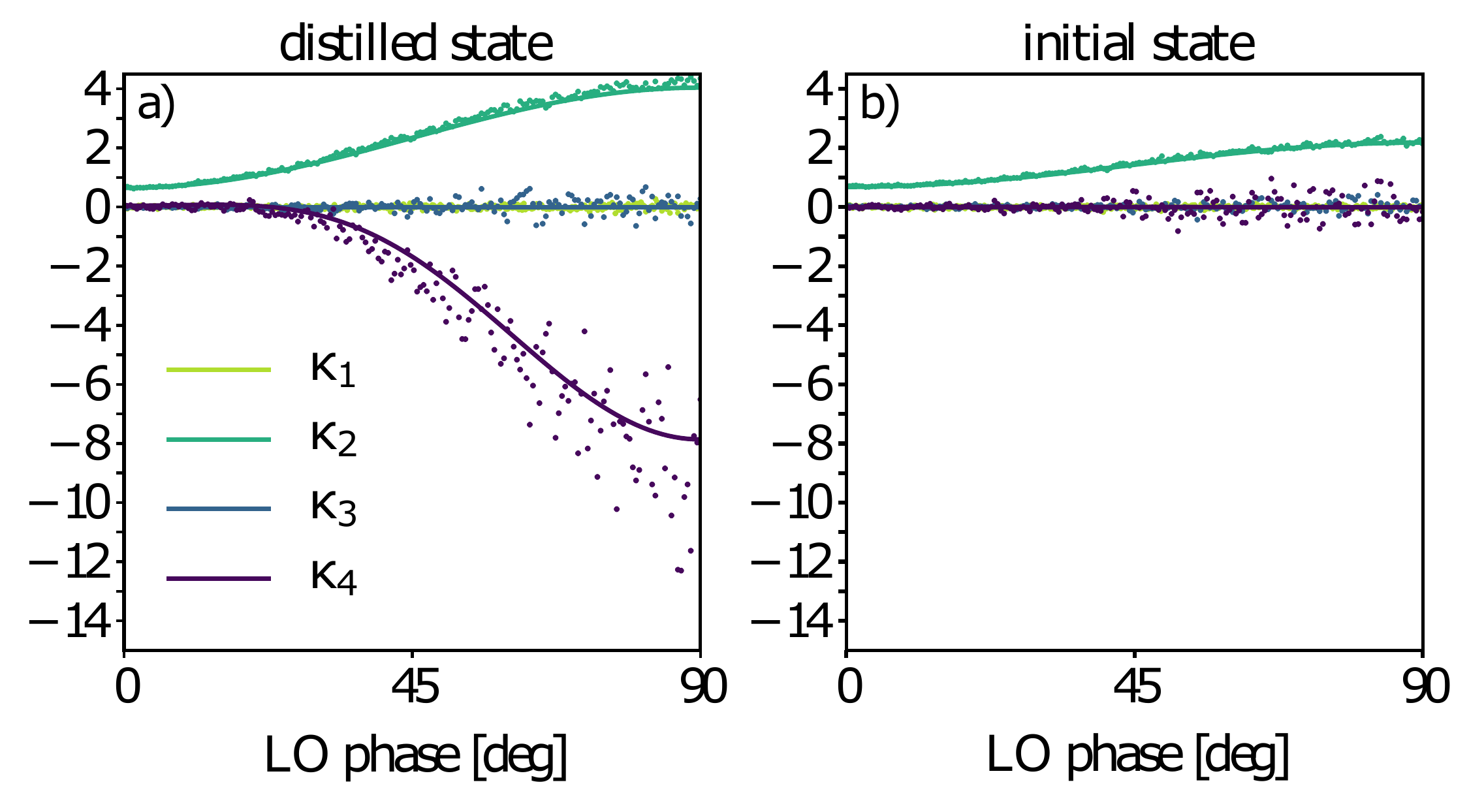}
\caption{Cumulants calculated from the raw data of the distilled a) and the initial squeezed state b) for different LO phases binned in steps of $0.5$ degrees. The dots indicate the values for the different cumulants calculated from the experimental data. The solid lines represent the expected values from calculating the cumulants using the simulated density matrices. The results are similar for both experimental and simulated cases: for the initial state, all cumulants except for the variance are centered around 0, as expected for a Gaussian state. The distilled state however shows a negative kurtosis with increasing phase, strongly hinting at its non-Gaussian nature. The strong spread in the values for the experimental data are visible due to their increased sensitivity to statistical fluctuations.}
\label{fig:cumulants}
\end{figure}

For our experimental realization of the distillation, no negativity of the Wigner function is expected even for the lossless case. To quantify the non-Gaussianity of the distillation process further, we investigate the higher-order cumulants \mbox{$\kappa_n, n \in \{1,2,3,4 \}$} of the Wigner function's marginal distribution as functions of the quadrature phase, where the projected quadrature is \mbox{$\hat{X}_{\theta} = \cos\theta \ \hat{X} +\sin\theta \ \hat{P}$} (see Fig.~\ref{fig:cumulants}). In terms of the central moments \mbox{$\mu_{n} = \left\langle \left( \hat{X}_{\theta} - \langle \hat{X}_{\theta} \rangle \right)^n\right\rangle$} and the mean value $m_1$, the cumulants can be expressed in a compact form as
\begin{eqnarray}
\kappa_1 &=& m_1,   \hspace{0.5cm} \kappa_2 = \mu_2 \nonumber \\
\kappa_3 &=& \mu_3, \hspace{0.5cm} \kappa_4 = \mu_4 - 3 \mu_2^2 .
\label{eq:Cumulants}
\end{eqnarray}
Gaussian distributions, such as the Wigner functions of pure squeezed states are fully determined by the first and second cumulant, i.e. by the mean value and the variance. Hence, particularly the higher-order cumulants characterize the peculiarities of the distilled state \cite{CRM2017}. The third and fourth cumulant are associated with the skewness and the 'tailedness' (kurtosis) of the distributions. For the undistilled state, we find that, apart from statistical fluctuations, only the variance $\kappa_{2}(\theta) = \mathrm{Var}(X_{\theta})$ is non-zero and shows the expected transition from squeezing $\mathrm{Var}(X_{\theta})<1$ to anti-squeezing $\mathrm{Var}(X_{\theta})>1$. For the distilled state, we find a similar behavior for $\kappa_{2}(\theta)$, however, with larger squeezing and anti-squeezing values. Moreover, we find that the kurtosis $\kappa_{4}(\theta)$ is non-zero and shows a clear trend to negative values when rotating the projected quadrature from the squeezing to the anti-squeezing angle.This feature is expected from our simulations as is indicated by the solid lines in Fig.~\ref{fig:cumulants}, which show the cumulants calculated from the simulated density matrices. For angles close to the squeezing quadrature, a small positive value is should be visible while a strictly negative kurtosis is calculated for the anti-squeezing quadrature. A negative kurtosis is a non-Gaussian feature showing that the distribution is flatter than a normal distribution. This fits well with the observed elongation of the Wigner function along the P quadrature in Fig.~\ref{fig:wignerfkt} which follows the prediction of our theoretical model.
\section{Conclusion}
In conclusion, we have demonstrated the realization of a pulsed squeezing distillation experiment using an engineered source of squeezed states. In our experiment, source engineering allows us to generate pulsed single-mode squeezed states to efficiently combine the mode-resolving homodyne detection technique and temporal-mode-insensitive photon counting measurements in a squeezing distillation protocol without the need of strong optical filtering. We demonstrate the expected increase in squeezing due to the distillation via photon subtraction and verify the non-Gaussian nature of the distilled quantum state by calculating the higher-order statistical cumulants. We provide a simple theoretical model describing the photon subtraction process in the few-mode regime and find an excellent agreement between the theoretical predictions and the experimentally found quadrature variances. Our model also shows, that the achieved improvement due to the distillation is limited due to technical reasons and the remaining mode impurity. It has been shown recently that it is possible to precisely manipulate the mode purity by using more elaborate means of spectrally shaping the pump field \cite{Ansari:18} or the utilization of a special poling pattern of the non-linear crystal \cite{Chen:19}. It is important to note, that the generated squeezing in this type of single-pass source is only limited by the amount of available pump energy. Experiments have reported the generation of states with a mean photon number $\langle \hat{n} \rangle = 80$ which translates to -25~dB of squeezing \cite{GeorgNIST}. Finally, optimizing the losses of optics and detectors in both the photon subtraction and the homodyne detection part of the setup will increase the achievable distillation yield and open up the possibility to subtract more photons per mode. Assuming perfect optics and detectors, and only considering the for our setup imperfect HOM visibility, an initially squeezed state with -3~dB of squeezing will, after subtracting two photons from each, the signal and idler mode, would exhibit a quadrature variance of -3.19 dB relative to shotnoise. All-together, the compatibility of our source with both homodyne and photon counting measurements renders it a versatile source of optical quantum states in hybrid QIP and a candidate for the realization of larger quantum networks requiring the interference of several similar sources. 
\section*{Funding}
This work has been funded by the European Union's Horizon 2020 research and innovation program through the Quantum-Flagship project CiViQ (no. 820466); JT, VA and CS acknowledge funding from European Union Horizon 2020 665148
(QCUMbER);
\section*{Disclosures}
The authors declare no conflict of interests.

\appendix
\section*{Appendix}
\section{Estimating the relative phase}
\label{app}
\begin{figure}[t]
	\centering
		\includegraphics[width=.65\columnwidth]{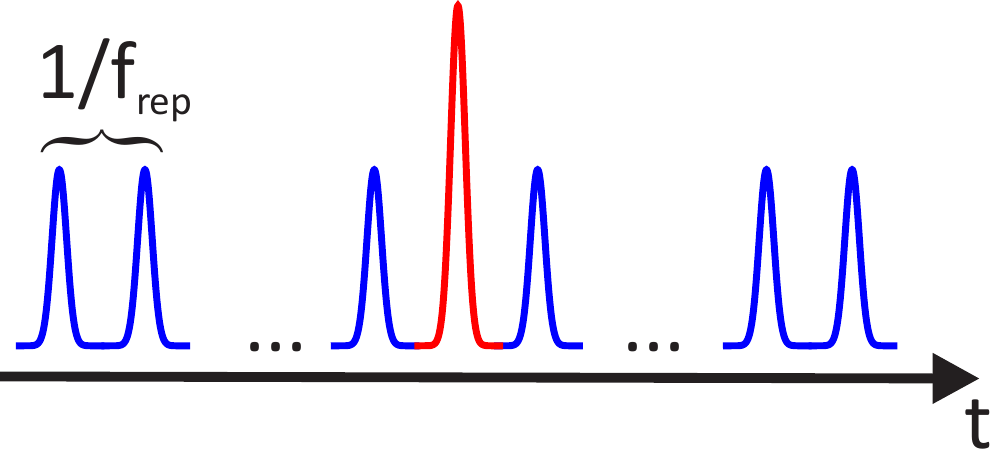}
	\caption{In each measured time trace, the distilled pulse (red) is surrounded by a several thousand undistilled, squeezed states (blue) which serve as reference to estimate the relative phase to the LO}
	\label{fig:supp_pulsetrain}
\end{figure}

 The main challenge in order to reconstruct the density matrix of the state is to assign the correct LO phase for each measurement. Often, this is achieved via the precise control of the relative phase between the LO and the detected state. In our case, we are unable to control the relative phase, thus we use the undistilled squeezed states surrounding each distilled pulse as a phase reference. We illustrate this situation in Fig.~\ref{fig:supp_pulsetrain}. This has the advantage that the overall experimental setup is simplified and that the whole phase assignment can be done in post-processing assuming that all phase angles are covered equally.
 \par
\begin{figure}[t]
	\centering
		\includegraphics[width=\columnwidth]{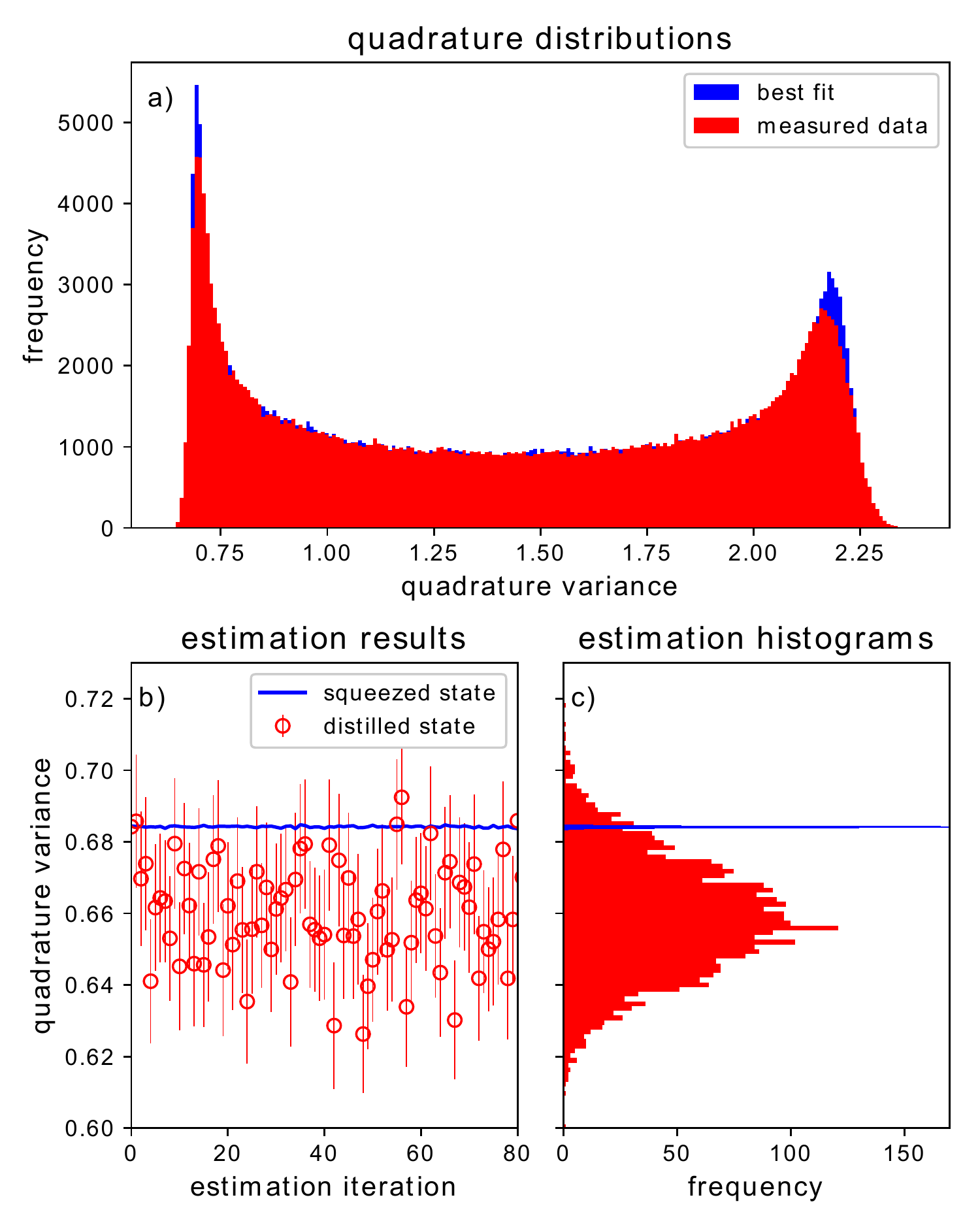}
	\caption{a) The distribution of the measured variances of the reference states (red) in comparison to the simulated distribution of the best fitting distribution of a squeezed state (blue); b) the probabilistic nature of the phase estimation leads to a variation of the variances, a shown here for the squeezing quadrature. Due to the higher number of constituting states, the spread of the squeezed state is smaller than for the distilled state; c) histograms of 3000 iterations of the phase estimation for squeezed and distilled states.}
	\label{fig:montecarlo}
\end{figure}

To ensure an even coverage of all possible phase angles, we therefore scan the LO phase using a piezo-mounted mirror with a modulation frequency in the range of $10~\mathrm{Hz}$. We choose the measurement window of our oscilloscope to encompass $100~\mathrm{\mu s}$, roughly consisting of 8000 pulses, as we assume that it is much faster compared to the phase shift introduced by the piezo modulation and the random phase fluctuations, e.g. vibrating mirrors.
In principle, different approaches to assign the LO phase are possible. We decided to use a MonteCarlo-like technique based on the experimental conditions to estimate the phase of each measurement which gives us a conservative estimate for the phase of each individual time trace of our measurement. This phase assignment procedure is described by the following steps:
\begin{enumerate}
	\item \textbf{Fitting the squeezing ellipse:} Assuming an equal sampling of all LO phases, we calculate the variance distribution by simulating the measurement of the marginal distribution \mbox{$\hat{X}_{\theta} = \hat{X}\cos\theta   +\hat{P}\sin\theta  $} of a squeezed state using $\hat{X}$ and $\hat{P}$ as fitting parameter. We then choose the combination of $\hat{X}$ and $\hat{P}$ matching the measured data best, see Fig.~\ref{fig:montecarlo}~a).
	\item \textbf{Phase-resolved simulation of the variances:} with the fitted squeezing and anti-squeezing values, we simulate homodyne measurements matching the experimental conditions, namely the detection of 8000 subsequent pulses and the calculation of their variance and the distribution under different phase angles $\mathrm{V}(\phi)$ assuming an uniform distribution of phases.
	\item \textbf{Inversion of the variance distribution:} the distribution found in the previous step is inverted into a variance-dependent distribution of phases $\phi(\mathrm{V})$.
	\item \textbf{Phase assignment for the individual measurement traces:} using the measured quadrature variances of the reference states, we randomly pick a possible phase from the phase distribution $\phi(\mathrm{V})$ to assign this measurement and the distilled pulse in its center a phase relative to the LO.
\end{enumerate}
However, our method is, at its core, still a probabilistic method, meaning that for different iterations one would receive different results. In Fig.~\ref{fig:montecarlo}b), we show the spread of the variance of the distilled state after 80 iterations of the sorting procedure using the same data together with the error bars indicating the statistical error. While the individually sorted data varies for each sorting outcome, the variances cluster around a common mean value, shown in Fig.~\ref{fig:montecarlo}c), which lies reasonably close to the value gathered from the simulation of the experiment. The mean of this spread is used to derive the value for the distilled squeezing of our experiment. However, we need to include it into the estimation of the experimental error, increasing it compared to the statistical error of the raw data alone.

\section{Loss estimation}
In our experiment, a multitude of individual parameters contribute to the overall loss figure. Most of them can be optimized in future experiments which would however not fit the scope of this first demonstration of source engineering in a hybrid QIP context. Most of the the individual contributions, such as the linear losses of the optics or the non-unity quantum efficiency of the photo diodes, can be remedied by further optimization. Other contributions cannot simply optimized by using different components. The HOM visibility of our source is intrinsically limited due to the phase matching conditions of KTP in this wavelength regime. Additionally, the the spatial overlap between LO mode and waveguide modes is limited due to differing geometrical profiles of the waveguide, which is close to quadratic, and a single-mode fiber, which is circular. Table~\ref{tab:2} lists the individual loss contributions of our setup and the theoretically possible squeezing of a squeezed state with an initial squeezing of -6 dB.
\begin{table}[t]\centering
\caption{\label{tab:2}Loss contributions of the experimental setup and the left-over squeezing of a squeezed state with 6 dB of initial squeezing.}
  \begin{tabular}{ccccc|c|c}
		\hline
HOM vis.& lin. losses& subtr. BS & LO vis. & PD quantum eff.& total eff. & squeezing [dB]\\
     \hline
0.75 & 0.87 & 0.9 & $(0.91)^2$ & 0.9 & 0.428 & -1.67\\
    \hline
  \end{tabular}
\end{table}

\clearpage
\bibliography{distillation_papers}
\end{document}